\documentclass[apjl]{emulateapj} 
\usepackage{apjfonts} 
 
\def\alwaysmath#1{\ifmmode{#1}\else{$#1$}\fi} 
 
\lefthead{Majewski et al.} 
\righthead{$\omega$Centauri Tidal Debris in the Grid Giant Star Survey}

\newcommand{\vgsr}{$v_{\rm GSR}~$}

\begin{document}

\title{Exploring Halo Substructure with Giant Stars:  
Substructure in the Local Halo as Seen in the Grid Giant Star Survey 
Including Extended Tidal Debris from $\omega$Centauri} 
 
\author{Steven R. Majewski\altaffilmark{1}, 
David L. Nidever\altaffilmark{1}, 
Verne V. Smith\altaffilmark{2}, 
Guillermo J. Damke\altaffilmark{1},\\
William E. Kunkel\altaffilmark{3}, 
Richard J. Patterson\altaffilmark{1}, 
Dmitry Bizyaev\altaffilmark{4},
and Ana E. Garc\'ia P\'erez\altaffilmark{1}
} 
 
\altaffiltext{1}{Dept. of Astronomy, University of Virginia, 
Charlottesville, VA 22904-4325 (srm4n, dln5q, gjd3r, ricky, aeg4x@virginia.edu)} 

\altaffiltext{2}{National Optical Astronomy Observatories, PO Box
26732, Tucson, AZ 85726 (vsmith@noao.edu)}
 
\altaffiltext{3}{Las Campanas Observatory, Casilla 601, La Serena,
  Chile (kunkel@jeito.lco.cl)}
  
\altaffiltext{4}{New Mexico State University/Apache Point Observatory, 
Sunspot NM 88349 (dmbiz@apo.nmsu.edu)}

\begin{abstract} 
 
We present the latitude-normalized radial velocity ($v_{b}$)
distribution of 3318 subsolar metallicity, $V$$\lesssim$$13.5$ stars
from the Grid Giant Star Survey (GGSS) in Southern Hemisphere fields.
The sample includes giants mostly within $\sim$5 kpc from the Galactic
disks and halo.  The nearby halo is found to (1) exhibit significant
kinematical substructure, and (2) be prominently represented by
several velocity coherent structures, including a very retrograde
``cloud" of stars at $l\sim285^{\circ}$ and extended, retrograde
``streams" visible as relatively tight $l$-$v_{b}$ sequences.  One
sequence in the fourth Galactic quadrant lies within the $l$-$v_{b}$
space expected to contain tidal debris from the ``star cluster"
$\omega$Centauri.  Not only does $\omega$Cen lie precisely in this
$l$-$v_{b}$ sequence, but the positions and $v_{b}$ of member stars
match those of $N$-body simulations of tidally disrupting dwarf
galaxies on orbits ending with $\omega$Cen's current position and
space motion.  But the ultimate proof that we have very likely found
extended parts of the $\omega$Cen tidal stream comes from echelle
spectroscopy of a subsample of the stars that reveals a very
particular chemical abundance signature known to occur only in
$\omega$Cen.  The newly discovered $\omega$Cen debris accounts for
almost {\it all} fourth Galactic quadrant retrograde stars in the
southern GGSS, which suggests $\omega$Cen is a dominant contributor of
retrograde giant stars in the inner Galaxy.
 
\end{abstract}

 
\section{Local Retrograde Halo Tidal Streams} 
 
If the Galactic halo has been formed partly
\citep{Searle1978} or entirely \citep[e.g.,][]{Majewski1993}
from accretion of smaller systems then one might expect groups of
stars with halo-like motions having strong velocity coherence
\citep[e.g.,][]{Helmi1999,Meza2005} near the Sun.  Claims for nearby
halo moving groups date back at least to \citet{Eggen1959} and include
several retrograde candidates -- including, in particular, the widely
recognized Kapteyn Group -- among the halo groups long discussed by
\citet[e.g.,][]{Eggen1965,Eggen1996a,Eggen1996b}.
\citet{Majewskietal1994,Majewskietal1996} analyzed nearby halo stars
(selected by asymmetric drift) towards the North Galactic Pole (NGP)
and claimed strong organization of their phase space distribution into
a few clumps, including a prominent retrograde moving group initially
identified via proper motions in \citet{Majewski1992}.  A possible
association of horizontal branch stars to this retrograde feature has
been identified by \citet{Kinman2007}.  \citet{Eggen1996b} suggests an
association of the \citet{Majewski1992} retrograde group to Kapteyn's
group.

More recently, interest in the notion that the globular cluster
$\omega$Centauri (``$\omega$Cen") may be the remnant core of a tidally
disrupted satellite galaxy \citep[e.g.,][]{Lee1999,Majewski2000b,Bekki2003} --- a notion inspired by (1) internal chemical and age
distributions belying multiple $\omega$Cen stellar populations, (2)
the example of the similarly massive cluster M54 located in/near/as
the ``core" of the disrupting Sagittarius galaxy \citep{Majewski2000b}
and apparent chemical similarities of M54+Sagittarius to $\omega$Cen
\citep{Carretta2010}, and (3) the unusual, low-inclination, retrograde
orbit of $\omega$Cen itself \citep{Dinescu2002} --- has led to several
$\omega$Cen tidal disruption simulations; these models generally
produce retrograde-moving, $\omega$Cen debris relatively near the
solar circle \citep{Dinescu2002,Tsuchiya2003,Tsuchiya2004,Mizutani2003,Bekki2003}.  This has prompted searches for retrograde
halo stars possibly shed by the ``cluster" and led to suggestions that
an ``$\omega$Cen" signal is present among local metal-weak stars
\citep{Dinescu2002,Mizutani2003,Meza2005}.  In fact,
\citet[e.g.,][]{Eggen1978} speculated a connection between his
Kapteyn's star group and $\omega$Cen \citep[see also][]{Kotoneva2005},
a dubious connection \citep{Proust1988} before the breadth of
$\omega$Cen's metallicity distribution function was fully recognized,
but more recently bolstered by detailed chemical analysis of Kapteyn
group stars \citep{Wylie2010}.

Despite these recent claims for stripped $\omega$Cen stars in the {\it
  solar neighborhood}, searches for extratidal stars near $\omega$Cen
itself have been less promising.  While the photometric search by
\citet{Leon2000} seemed to suggest a ``significant" pair of tidal
tails extending from $\omega$Cen, these results were cast in doubt
when the substantial foreground differential reddening was assessed
\citep{Law2003}.\footnote{\citet{Leon2000} themselves warned that dust
  extinction might be influencing their results.}  An expansive
spectroscopic search for $\omega$Cen stars beyond its tidal radius by
\citet{DaCosta2008} reveals only six candidates among more
than 4,000 stars selected from the $\omega$Cen giant branch in the
color-magnitude diagram; these authors suggest that this meager haul
is consistent with models where most stripping took place long ago,
and with the lost stars now widely distributed about the Galaxy.
Stronger support for the tidally-disrupted dwarf galaxy model, and for
confidently linking solar neighborhood candidate members with
$\omega$Cen itself, would come from actually being able to trace
debris along the satellite's orbit, and, eventually, from $\omega$Cen
continuously to the solar neighborhood.

Here we report detection of a kinematically coherent ``tidal debris"
signature spanning $\gtrsim$$60^{\circ}$ of Galactic longitude in a
large radial velocity (RV) survey of giant stars mostly ($\sim95\%$)
within $\sim$$5$ kpc of the Sun.  Stars within this dynamically
coherent group show a specific chemical marker thought to be unique to
$\omega$Cen, as well as distances and velocities consistent with
models of $\omega$Cen tidal debris.  Though still mostly only a few
kiloparsecs away, these extended tidal debris stars provide a start at
tracing the $\omega$Cen stream and provide crucial, though still
crude, dynamical constraints on the disruption of the closest known
dwarf galaxy to the Sun.  At minimum, these discovered debris stars
suggest that $\omega$Cen may be a principal source of local retrograde
stars.

\section{Substructure in the Southern GGSS}

Our analysis uses the Grid Giant Star Survey (GGSS), a
partially-filled, all-sky search for giant stars using the Washington
$M,T_2$+$DDO51$ photometric selection technique described in
\citet{PaperI}.  The GGSS had as one goal the
identification of bright but distant, metal-poor giants suitable for
the Astrometric Grid of the (now defunct) Space Interferometry
Mission.  To this end, a specific subset of potential SIM Grid stars
was drawn from the GGSS: the four most distant giant stars with
$M<13.5$ in each of the 1302 evenly spaced, $\sim0.4$-$0.6$ deg$^2$
GGSS fields, where photometric distances were estimated by assigning
absolute magnitudes to stars from their position in the
($M-T_2,M-DDO51$) diagram \citep[which separates dwarf and giant stars and,
within these luminosity classes, sorts by metallicity;][]{PaperI}.  This
particular selection biases this ``SIM Grid sample" to more metal poor
giants, because these are farther at a given apparent magnitude.  An
echelle resolution study of 774 Grid candidates selected in this way
verified 100\% of them to have giant branch surface gravities, and
indicated a median [Fe/H]$\sim$$-0.7$ and distance of $\sim$2 kpc
\citep{Bizyaev2006}, but with long tails to more distant and
metal-poor stars.  Thus, the sample contains a mix of old thin disk,
Intermediate Population II (IPII) thick disk and a smaller fraction of
halo stars in mean proportions varying by Galactic latitude.  Our
initial analysis here relies on $R$$\sim$$2,600$ spectroscopy of a
larger sample of 3318 GGSS ``SIM Grid" giants from our southern
observing campaign (sky distribution shown in Fig.\ 1a); further
discussion of the GGSS is given in \citet{Patterson2001} and
S. Majewski (in preparation).
 
In 1999 a spectroscopic follow-up campaign for these candidate
Astrometric Grid stars began at Las Campanas Observatory using the
Modular Spectrograph on the Swope 1-m telescope (and the DuPont 2.5-m
for 68 stars).  The typical set-up used the 600 line/mm grating to
sample a $\sim2000$\AA\ spectral region including both H$\alpha$ and
H$\beta$ at 1.0\AA\ pixel$^{-1}$ (in a small fraction of cases the Ca
infrared triplet region was observed; details of both setups are in
\citet{Majewskietal2004}.  The $S/N$ of the spectra exceeds 10 in almost
all cases and reaches $>$$50$ in some cases.  RV cross-correlation of
the spectra used the methodology discussed in \citet{Majewskietal2004}.
Typical RV errors are 5-10 km~s$^{-1}$, estimated a variety of ways
including repeat measures of many stars and comparison to echelle
resolution RVs.

\begin{figure}[th]
\epsscale{0.70}
\includegraphics[angle=0,scale=0.42]{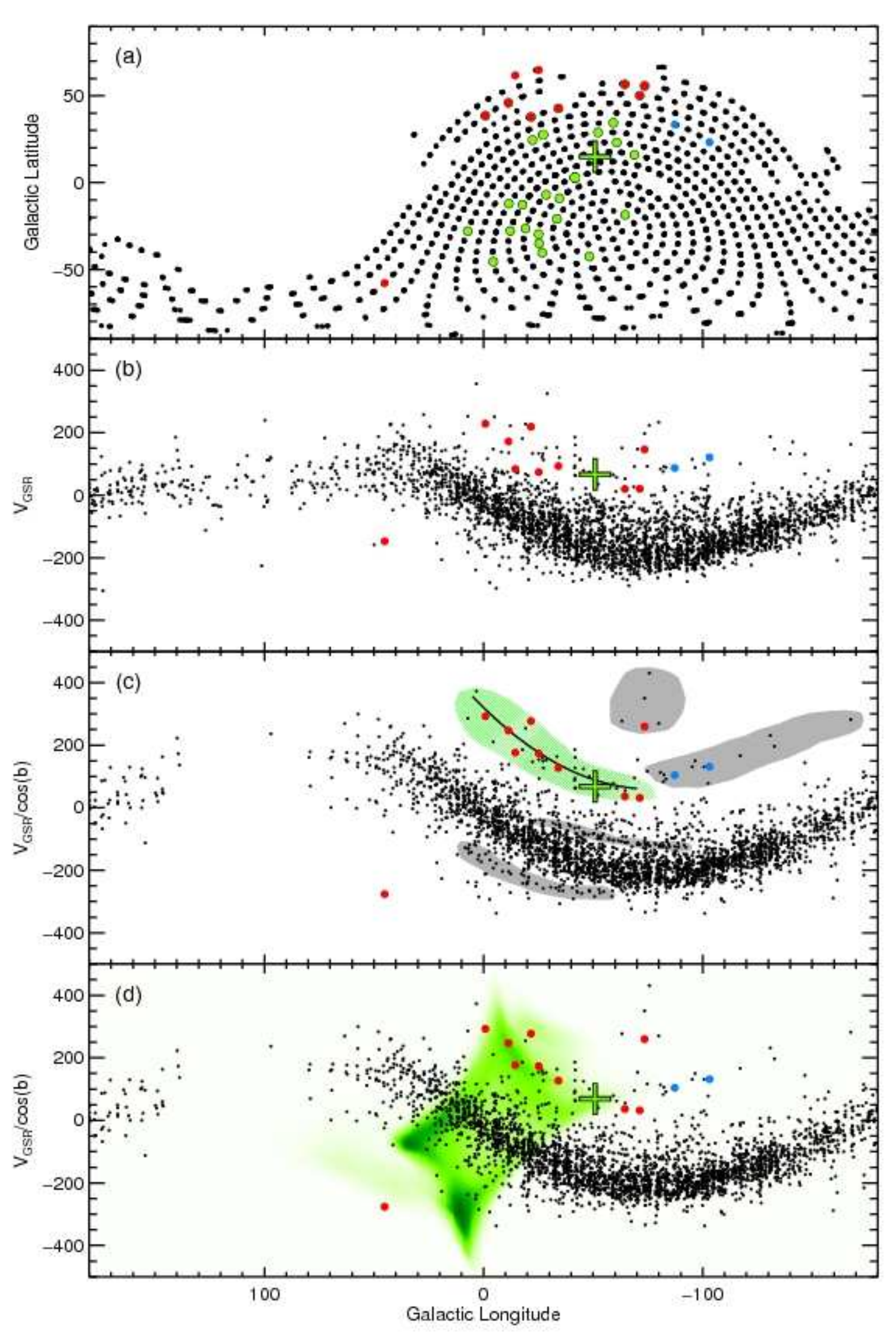}
\caption{The distribution of the GGSS giant stars used in this study
  in (a) Galactic coordinates, (b) \vgsr versus longitude, and (c)
  $v_{b}$ = \vgsr $/ \cos{b}$ versus longitude (for only $|b|<$60\degr
  in panel c).  The curve and green shading in (c) highlight the arc
  of stars we believe contains $\omega$Cen debris.  Stars in this
  sequence are marked with green or red points in panel (a).  Grey
  shadings show several other potential halo substructures.  The
  position of the $\omega$Cen core in all panels is shown by the large
  green cross.  Panel (d) is the same as panel (c) but showing the
  ``probability distribution" of $\omega$Cen tidal debris based on our
  suite of models.  Superposed red and blue points in all panels
  represent stars having echelle spectra with red designating those
  stars that follow the $\omega$Cen [Ba/Fe]-[Fe/H] patterns and blue
  those that do not.}
\end{figure}

Figure 1b shows the distribution of Galactic Standard of Rest (GSR)
RVs ($v_{GSR}$) for the ``SIM Grid" stars, assuming a solar motion in
right-handed Galactic coordinates of $(+10.0,+225.3,+7.2)$
km~s$^{-1}$.  For stars moving predominantly in Galactic planar orbits
(e.g., disk stars {\it and} putative $\omega$Cen debris), the
approximate ``planar RV" (i.e., the observed RV the star would have
were it on the Galactic equator) is given by $v_b=v_{GSR}/\cos(b)$
(this latitude normalization breaks down at high latitude, where a
star's $Z$ motion dominates the observed RV, so Fig.\ 1c is limited to
stars with $|b|<60^{\circ}$).  In this planar projection stellar
populations more clearly sort by their relative asymmetric drifts, and
Figure 1c for the most part shows the expected longitudinal
distribution of $v_{GSR}$ for a predominantly thin disk/IPII mix of
stars.
 
However, stars not following disk kinematics are evident, including a
number having retrograde velocities.  Among stars with halo-like
velocities (and particularly among likely retrograde stars), outlier
groupings (e.g., at [$l$, $v_{b}$] $\sim$ [$280^{\circ}$, 300
  km~s$^{-1}$]) or thin, coherent strands of stars (e.g., from
[$20^{\circ}$, $-125$ km~s$^{-1}$] to [$300^{\circ}$, $-275$
  km~s$^{-1}$]) can be seen in Figures 1b and/or 1c (where we have
highlighted some interesting features with shading).  Such cold and
coherent RV trends with sky position are characteristic of long tidal
streams, such as the Sagittarius system
\citep[e.g.,][]{Majewskietal2004, Law2005}; in this case, however, the
substructure is found among relatively nearby giants and therefore
corresponds to stars with a much broader sky distribution than typical
for more distant streams (Fig.\ 1a).  Such substructure among the GGSS
giants is not surprising given that they probe distances similar to
the mostly main sequence stars in the
\citet{Majewskietal1994,Majewskietal1996} study, which also showed
significant halo substructure (but in only a single pencil beam).
Moreover, a new ``all-sky" study of bright M giants by A. Sheffield
(in preparation) and probing comparable distances shows analogous,
though even more striking, [$l$, $v_{b}$] coherences among halo-like
stars (most likely because M giants probe typically younger tidal
streams).  Together, the GGSS giants, Sheffield M giants,
\citet{Majewskietal1994,Majewskietal1996} subdwarfs, and
\citeauthor{Kinman2007} horizontal branch stars point to the high
degree of substructure in the halo even at the solar circle.  Indeed,
the degree of velocity coherence and substructure of the local halo
does not differ much from that seen in the distant halo
\citep{Majewski2004}.

\section{Tidal Debris Model and the $\omega$Centauri Connection}

Are there stars in the GGSS sample that can be associated with
$\omega$Cen?  The studies of solar neighborhood $\omega$Cen debris
mentioned in \S1 were able to make use of the expected ($E, L_z$)
distributions to trawl for the best local representatives.
Unfortunately, our giant stars are far enough away that most available
proper motions (and therefore complete space velocity determinations)
are unreliable.  Therefore we winnow our search to those [$l$,
  $v_{b}$] ranges expected to be populated by any realistic model of
$\omega$Cen tidal disruption.  To do so we create a suite of N-body
simulations of satellites undergoing tidal disruption along
$\omega$Cen-like orbits in the static Milky Way (MW) potential given
by \citet{Johnston1995}, with 30,000 particles representing the
parent satellite in an initially Plummer configuration.  To account
for observational uncertainties that prohibit us from accurately
knowing the true position, space motion, and therefore orbits of both
$\omega$Cen and the Sun, we create a grid of models spanning ranges of
uncertainty around typical mean values for each critical interaction
parameter \citep[from][]{Dinescu1999}: $\omega$Cen distances of
[4.9,5.1,5.3] kpc, Galactic Cartesian velocities (right-handed system
in the Galactic rest frame) of $V_x$ = $[42,53,64,75,86]$ km~s$^{-1}$,
$V_y$ = $[-52,-43,-34,-25,-16]$ km~s$^{-1}$, and $V_z$ = $[-6,4,14]$
km~s$^{-1}$, a solar Galactocentric distance of $[7.0, 7.75, 8.5]$
kpc, and, to set the MW mass scale, a local circular velocity of
$[220, 254]$ km~s$^{-1}$, where the latter value is that suggested by
\citet{Reid2009}.  Within this grid we adopt satellites of three
different initial masses, [3e7, 3e8, 3e9] M$_{\odot}$, and evolved for
$\sim0.4$ Gyr, ensuring that each model orbit places the satellite at
the current ($l$,$b$) position and RV of $\omega$Cen.

These 4050 simulations produce a variety of tidal streams from which
we can establish the range of possible [$l$, $v_{b}$] distributions;
in fact, the sum of these models (Fig.~1d) gives us something like a
``probability distribution function" (PDF) of where the last 0.4 Gyr
of $\omega$Cen tidal debris might most likely lie (Fig.\ 1d).
Comparing this to the GGSS distribution, and ignoring regions
dominated by disk stars, we call attention to a particularly striking
match of the PDF to a relatively tight sequence of likely retrograde
stars over $l \sim280$-$360^{\circ}$, highlighted with green shading
in Figure 1c.  That these stars form a coherent velocity structure
orbiting in a near-Galactic planar orbit is demonstrated by the fact
that, despite their broad sky distribution (green and red points,
Fig.\ 1a), they show a coherent, string-like configuration in planar,
$v_{b}$ projection.

A 2nd-order polynomial fit to the Figure 1c sequence with iterative
rejection settles on 35 stars in the feature with an observed RV
dispersion of only $\sim$40 km~s$^{-1}$.  Most interestingly, the
sequence passes through the ($l$, $v_{b}$) position of $\omega$Cen,
also shown.

\begin{figure}[t]
\includegraphics[angle=0,scale=0.42]{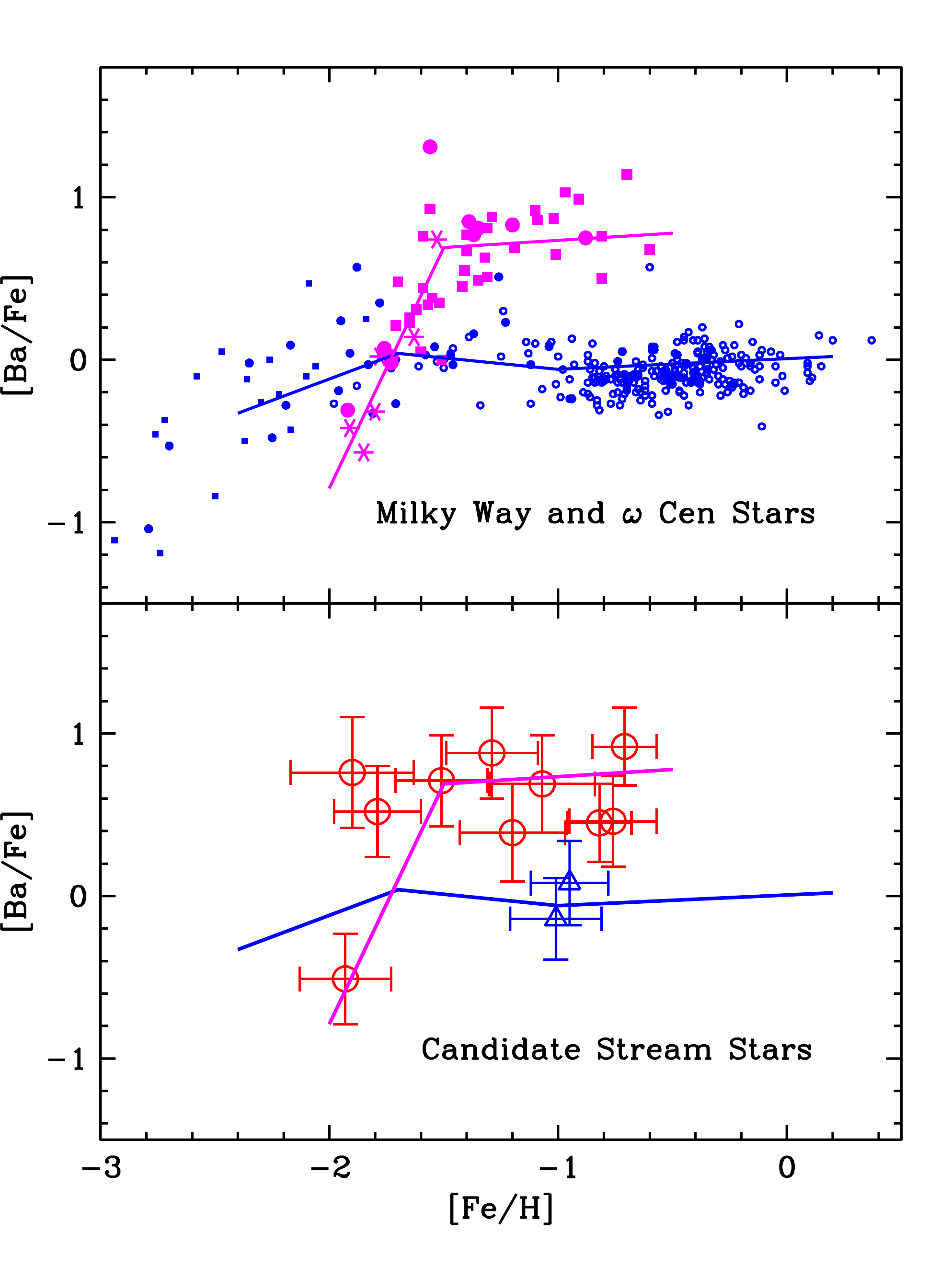}
\caption{(Top panel) The distribution of [Ba/Fe]-[Fe/H] for MW stars
  (blue points) and a characteristic locus (blue line) from data by
  \citet{Fulbright2002}, \citet{Johnson2002}, and \citet{Reddy2003,Reddy2006},
  overlaid with the same for $\omega$Cen stars (magenta points and
  line) from \citet{Francois1988}, \citet{Norris1995}, and
  \citet{Smith2000}.  (Bottom panel) The distribution of barium
  abundances for the ten stars following the retrograde sequence lying
  within the ``$\omega$Cen PDF region" and containing the position of
  $\omega$Cen shown in Fig. 1 (red points) versus those stars lying
  outside the ``$\omega$Cen PDF region" (blue).  The colored lines are
  those shown in the top panel.}
\end{figure}

To test whether $\omega$Cen stars might fall among those GGSS stars
lying within this retrograde feature, high-resolution
($R$$\sim$55,000), echelle spectra for eight of them, as well as a
control sample of four other halo GGSS stars (including two very
extreme $v_{b}$ stars at similar longitudes), were obtained using the
Sandiford echelle spectrometer \citep{McCarthy1993} on the McDonald
2.1-m Struve telescope.  Given the modest wavelength coverage and
$S/N$$\sim$25-50 of the spectra, iron abundances were derived from a
set of unblended \ion{Fe}{1} lines using measured equivalent widths.  The
stellar parameters $T_{\rm eff}$ and $\log{g}$ were taken from the
analysis of GGSS stars by \citet{Bizyaev2006}, with determinations
of the microturbulence velocities ($\xi$) set by the \ion{Fe}{1} lines
measured for this study.

The well-defined \ion{Ba}{2} line at 5854\AA\ was used as an s-process
abundance indicator because (1) it is the most easily detectable in
these generally mediocre spectra, and (2) the distribution of [Ba/Fe]
-- [Fe/H] for $\omega$Cen is quite distinctive (Fig.~2a) --- indeed it
is unique among all star systems studied to date in the extreme
overabundances of s-process elements characterizing its more
metal-rich stellar population (see Fig.~11 of the Geisler et al.~2007 review).
Figure 2b shows the derived [Ba/Fe]--[Fe/H] pattern for the
12 GGSS stars; abundances and other data derived for these stars are
given in Table 1.  The quoted velocities are from the original
medium-resolution spectra and have typical random errors of $\sim$10
km~s$^{-1}$; photometric uncertainties are about 0.01 mag.  Abundance
uncertainties (shown in Fig. 2b) were set by the sensitivities of \ion{Fe}{1}
and \ion{Ba}{2} abundances to changes in stellar parameters of $\pm$100K in
$T_{\rm eff}$, $\pm$0.3 dex in $\log{g}$, and $\pm$0.5 km~s$^{-1}$ in
$\xi$.  As is vividly demonstrated, the Table 1 stars most likely to
be kinematically associated with $\omega$Cen (Fig.\ 1d) clearly follow
the characteristic $\omega$Cen trend in [Ba/Fe] versus [Fe/H], while
two stars least likely to be kinematically associated with $\omega$Cen
lie along the MW trend.  The combination of kinematical consistency
and possession of the hallmark barium abundance trends for the former
group of stars are strong evidence that an extended part of the
$\omega$Cen tidal debris stream has been found.\footnote{The two Table
  1 stars with $\omega$Cen chemistry and extreme $v_b$, off the main
  ``green trend" in Fig. 1c, are plausibly associated with older
  $\omega$Cen tidal debris wraps, shown with faint probability in
  Fig.\ 1d, or not currently part of our 0.4 Gyr-long models (see
  Fig.\ 4).}

It is worth noting that within this rather small sample of $\omega$Cen
stream candidates is a carbon-rich star, G1358-16.167.  This red giant
exhibits strong Swan C$_{\rm 2}$ bands (e.g., at $\lambda$5165\AA;
Fig.\ 3) and is strongly barium-enhanced.  Such C-rich halo giants at
this modestly low metallicity are relatively rare and constitute only
about 1-2\% of halo giants.  However, $\omega$Cen has at least five
known C-rich giants \citep[see Table 3 in][which both subgiants and
  giants]{Bartkevicius1996} whereas the only other globular clusters
known to harbor C-rich giants are M22 (with two --- as well as a
spread in heavy-element abundances and s-process enrichment, similar
to, but not to the degree of, $\omega$Cen), M2 (with one), and M55
(with one --- \citet*{Smith1982}).  Given the likely association of
most Table 1 stars to $\omega$Cen it is not too surprising that one is
found to be C-rich; this observation only strengthens the tie to their
chemically peculiar parent stellar system.

\begin{figure}[t]
\includegraphics[angle=0,scale=0.34]{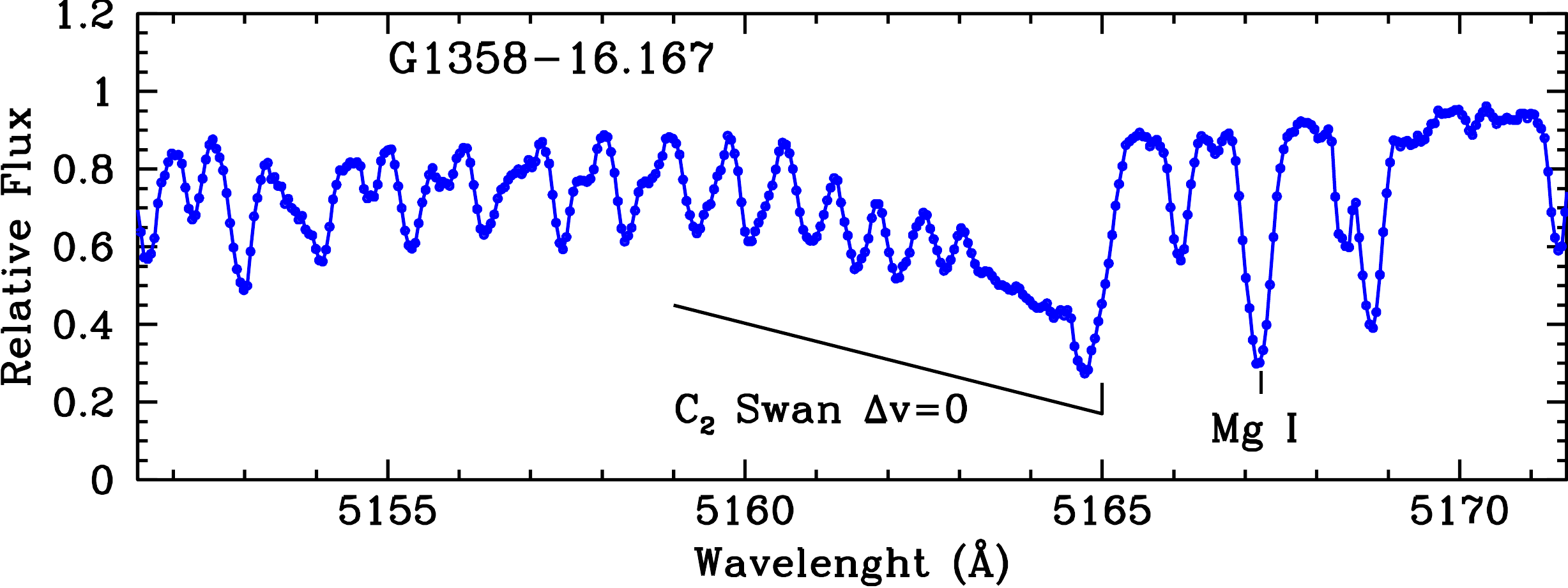}
\caption{Example of a C$_{\rm 2}$ Swan band in the spectrum of G1358-16.167.}
\end{figure}

Based on the [Fe/H]-[$\alpha$/Fe] and age-metallicity distributions
for $\omega$Cen given by \citet{Stanford2006} --- adopting a 4 Gyr
$\omega$Cen age span --- and the derived [Fe/H] and 2MASS photometric
data for the ten good GGSS $\omega$Cen candidates we estimate their
distances using matching \citet{Dotter2008} isochrones.  With these
distances we can place the stars in their Galactic planar positions
relative to $\omega$Cen and the Sun (Fig.\ 4).  Superposed on this
distribution we plot the model debris and satellite orbit from one of
several $\omega$Cen models in our grid (the example model parameters
are given in the figure legend) that provide a reasonable match to the
positions and RVs of the GGSS stars.  This model, based
on a satellite orbit with peri-/apo-Galactica limits of (1 kpc)/(7
kpc), respectively, not only demonstrates how the stars of interest
very plausibly trace $\omega$Cen debris, but also how it might be
possible for $\omega$Cen tidal debris to reach the solar neighborhood,
as suggested by the various claims for this discussed in \S1.

In fact, though, as stated earlier, available proper motions for the
GGSS $\omega$Cen stars are typically of low quality, the UCAC
astrometry \citep{Zacharias2010} does hint at further tantalizing
connections to previous claims for nearby $\omega$Cen debris: The GGSS
$\omega$Cen stars with the smallest derived ($E, L_z$) uncertainties
happen to fall in, or quite near, the ``$\omega$Cen debris
expectation" box defined in ($E, L_z$) by \citet[][using the same
  gravitational potential]{Dinescu2002}, whereas the weighted mean $L_z$ of all
ten GGSS $\omega$Cen stars, $-179\pm135$ km~s$^{-1}$ kpc, matches
extremely well the ``$\omega$Cen peak" identified within the stellar
sample explored by \citet[][their Fig.\ 9]{Meza2005}.

\begin{figure}[t]
\includegraphics[angle=0,scale=0.35]{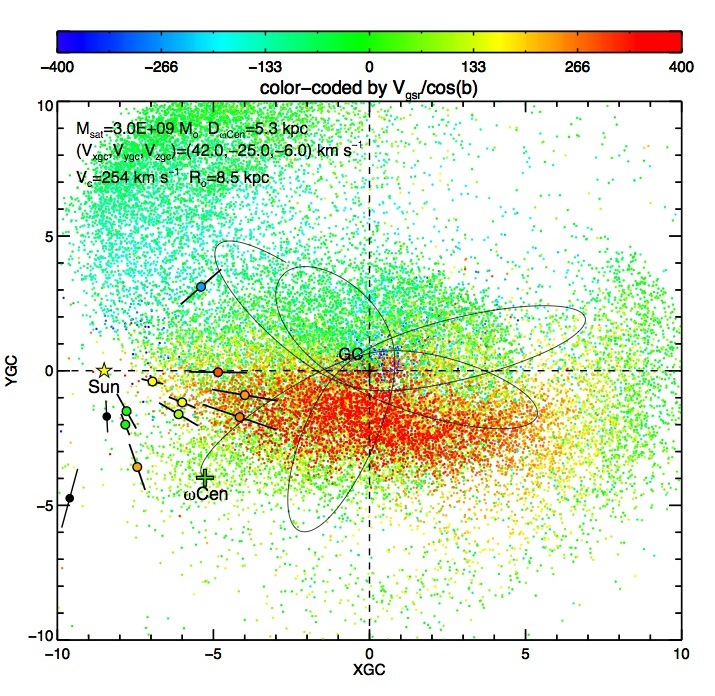}
\caption{The Galactic X-Y positions of Table 1 stars with respect to
  the Sun ($\odot$) and $\omega$Cen (large green cross).  Overlaid is
  an N-body simulation of $\omega$Cen tidal disruption and the
  associated orbit over the past 0.4 Gyr with the parameters given in
  the legend.  The model and Table 1 stars with $\omega$Cen-like
  chemistry are color-coded by $v_b$, which match well in general at
  the computed positions of the Table 1 stars. The two black points
  are the Table 1 stars that appear to be kinematically, spatially and
  chemically unassociated with $\omega$Cen.}
\end{figure}

\section{Some Implications} 

In the GGSS sample of bright giant stars within $\sim$5 kpc across the
southern 2/3 of the celestial sphere and biased toward metal-poor
stars we have shown evidence that those having halo-like velocities,
and particularly those with retrograde velocities, show a highly
substructured, rather than random, distribution.  This result is a
further demonstration that even in the inner halo, at the position of
the solar circle, the Galactic halo is not well-mixed, but shows the
signature of multiple minor accretion events.  We identify one group
of stars kinematically and chemically consistent with being
$\omega$Cen debris, which we use to get a rough constraint on the
$\omega$Cen tidal stream in the inner Galaxy, a model demonstrating
how $\omega$Cen stars can reach the solar neighborhood.  We have
probed with high resolution spectroscopy less than half of the several
dozen stars we associate with $\omega$Cen [$l$,$v_{b}$] features, yet
all of the best candidates are found to be chemically consistent with
being from $\omega$Cen; we can project that most likely a large
fraction of the fuller sample of ``$\omega$Cen" candidates are
authentic.  Considering that only a few dozen clearly retrograde stars
are found in the southern GGSS at all (Fig.\ 1b-c), our results
suggest that $\omega$Cen tidal debris is a primary contributor of
retrograde stars near the Sun, and likely the overwhelmingly dominant
contributor in the inner two Galactic quadrants.

\acknowledgements We gratefully acknowledge support by NASA/JPL
contracts 1201670 and 1228235, and the David and Lucile Packard
Foundation to build the GGSS, as well as NSF grants AST-0307851 and
AST-0807945.  This work would not have been possible without the
generous access to the Swope telescope for development of the SIM
Astrometric Grid granted by Carnegie Observatories Directors Augustus
Oemler and Wendy Freedman.

\newpage

\begin{deluxetable}{lccrrccrcc@{~}cl@{$\pm$}ll@{$\pm$}lrl@{$\pm$}l}
\tablecolumns{18}
\tabletypesize{\scriptsize}
\tablewidth{0pt}
\setlength{\tabcolsep}{0.05in}
\tablecaption{GGSS Stars with Echelle Spectroscopy}
\tablehead{
\colhead{Star}&  \colhead{RA}   &\colhead{Dec} &
\colhead{$l$}& \colhead{$b$}& \colhead{$M$}& \colhead{$M$$-$$T_2$} & \colhead{$V_{\rm GSR}$} &   \colhead{T$_{\rm eff}$} &
\multicolumn{2}{l}{{$\log{g}$}~~~~~~{$\xi$}}  & \multicolumn{2}{c}{[Fe/H]}  &\multicolumn{2}{c}{[Ba/Fe]} &  \colhead{Age\tablenotemark{a}} & \multicolumn{2}{c}{$D_{\rm 2MASS}$}  \\
\colhead{} &  \multicolumn{2}{c}{(J2000.0)} & \colhead{(${}^{\circ}$)} & \colhead{(${}^{\circ}$)}&\colhead{}&\colhead{} &\colhead{(km~s$^{-1}$)}
& \colhead{(K)} &\colhead{} & \colhead{~~(km~s$^{-1}$)} & \multicolumn{2}{c}{(dex)} & 
 \multicolumn{2}{c}{(dex)}&\colhead{(Gyr)} & \multicolumn{2}{c}{(kpc)} 
} \\
\startdata
\sidehead{\it Stars least likely associated with $\omega$ Centauri}
   G0945$-$22.153 & 09:47:51.6&$-$22:48:44 &257&  23 & 13.19& 1.56&     121.0  &   4200&   1.5 &  2.0 &$-0.95$&    0.17&$ +0.08$&    0.26& 10.8 &  6.1&  1.3\\
   G1100$-$22.138 & 11:03:06.6&$-$22:57:50 &273&  33 & 13.23& 1.29&      86.9  &   4800&   1.9 &  3.0 &$-1.01$&    0.15&$ -0.14$&    0.25& 11.0 &  2.8&  0.7\\
  
\sidehead{\it Stars most likely associated with $\omega$ Centauri}  
   G1211$-$05.126 & 12:14:34.6&$-$05:57:53 &287&  56 & 13.22& 1.81&     146.3  &   4075&   1.9 &  2.1 &$-1.07$&    0.23&$ +0.69$&    0.30& 11.2 &  7.3&  1.6\\
   G1211$-$11.146 & 12:14:23.6&$-$11:43:49 &289&  50 & 12.52& 1.58&      20.4  &   4250&   2.2 &  1.4 &$-0.82$&    0.14&$ +0.45$&    0.24& 10.3 &  3.7&  0.7\\
   G1232$-$05.60  & 12:35:00.3&$-$06:02:30 &296&  57 & 12.60& 1.17&      23.7  &   4800&   1.1 &  2.5 &$-1.93$&    0.20&$ -0.51$&    0.28& 13.5 &  3.5&  1.0\\
   G1341$+$05.29  & 13:43:31.1& $+$05:15:39 &335&  65 & 13.45& 1.73&      73.8  &   4000&   2.0 &  2.0 &$-0.71$&    0.14&$ +0.92$&    0.24&  9.9 &  6.8&  1.1\\
   G1358$-$16.167 & 14:01:01.4&$-$17:00:24 &326&  43 & 12.55& 1.47&      93.4  &   4400&   1.9 &  1.8 &$-1.29$&    0.20&$ +0.88$&    0.28& 12.0 &  5.0&  1.1\\
   G1403$+$05.80  & 14:06:34.1& $+$05:12:41 &346&  62 & 12.86& 1.43&      83.8  &   4375&   2.2 &  2.2 &$-0.76$&    0.19&$ +0.46$&    0.28& 10.1 &  3.5&  0.8\\
   G1443$-$16.11  & 14:45:52.9&$-$17:03:06 &339&  38 & 12.65& 1.70&     219.0  &   4375&   1.7 &  1.5 &$-1.90$&    0.27&$ +0.76$&    0.34& 13.5 &  8.0&  1.9\\
   G1448$-$05.103 & 14:51:32.3&$-$06:00:04 &349&  46 & 13.30& 1.65&     171.9  &   4250&   2.1 &  2.4 &$-1.20$&    0.23&$ +0.39$&    0.30& 11.7 &  7.8&  1.7\\
   G1532$-$05.29  & 15:34:27.9&$-$05:52:49 &  0&  39 & 13.12& 1.52&     228.8  &   4500&   2.6 &  1.7 &$-1.51$&    0.20&$ +0.71$&    0.28& 12.8 &  6.7&  1.5\\
   G2237$-$16.2017& 22:39:41.9&$-$16:35:46 & 45&$-58$& 12.49& 1.60&    $-146.9$&   4150&   1.9 &  2.4 &$-1.79$&    0.19&$ +0.52$&    0.28& 13.5 &  9.2&  1.6 
\enddata
\tablenotetext{a}{Assumed age based on the $\omega$Cen age-metallicity relation by \citet{Stanford2006} and our derived [Fe/H].}
\end{deluxetable}


\begin{thebibliography}{} 
 
\bibitem[Bartkevicius(1996)]{Bartkevicius1996} Bartkevicius, A. 1996,
  Baltic Astronomy, 5, 217

\bibitem[Bekki \& Freeman(2003)]{Bekki2003} Bekki, K. \& Freeman,
  K.~C.\ 2003, \mnras, 346, L11
 
\bibitem[Bizyaev et al.(2006)]{Bizyaev2006} Bizyaev, D., et al.\ 2006,
  \aj, 131, 1784

\bibitem[Carretta et al.(2010)]{Carretta2010} Carretta, E., et
  al.\ 2010, \apjl, 714, L7

\bibitem[Da Costa \& Coleman(2008)]{DaCosta2008} Da Costa, G.~S. \&
  Coleman, M.~G.\ 2008, \aj, 136, 506

\bibitem[Dinescu(2002)]{Dinescu2002} Dinescu, D.~I.\ 2002, ASP
  Conf.~Ser.~265: Omega Centauri, A Unique Window into Astrophysics,
  265, 365
 
\bibitem[Dinescu et al.(1999)]{Dinescu1999} Dinescu, D.~I., Girard,
  T.~M., \& van Altena, W.~F.\ 1999, \aj, 117, 1792

\bibitem[Dotter et al.(2008)]{Dotter2008} Dotter, A., Chaboyer, B.,
  Jevremovi{\'c}, D., Kostov, V., Baron, E., \& Ferguson, J.~W.\ 2008,
  \apjs, 178, 89

\bibitem[Eggen(1965)]{Eggen1965} Eggen, O. J. 1965, in Galactic
  Structure, eds. A. Blaauw \& M. Schmidt, p. 111

\bibitem[Eggen(1978)]{Eggen1978} Eggen, O. J. 1978, \apj, 221, 881

\bibitem[Eggen(1996a)]{Eggen1996a} Eggen, O. J. 1996a, \aj, 112, 1595

\bibitem[Eggen(1996b)]{Eggen1996b} Eggen, O. J. 1996b, \aj, 112, 2661

\bibitem[Eggen \& Sandage(1959)]{Eggen1959} Eggen, O. J. \& Sandage,
  A.~R.\ 1959, \mnras, 119, 255

\bibitem[Francois et al.(1988)]{Francois1988} Francois, P., Spite, M.,
  Spite, F. 1988, \aap, 191, 267

\bibitem[Fulbright(2002)]{Fulbright2002} Fulbright, J.~P. 2002, \aj,
  123, 404

\bibitem[Geisler et al.(2007)]{Geisler2007} Geisler, D., Wallerstein,
  G., Smith, V.~V., Casetti-Dinescu, D.~I. 2007, \pasp, 119, 939
 
\bibitem[Helmi \& White(1999)]{Helmi1999} Helmi, A. \& White,
  S.~D.~M.\ 1999, \mnras, 307, 495

\bibitem[Johnson(2002)]{Johnson2002} Johnson, J. A. 2002, \apjs, 139,
  219

\bibitem[Johnston et al.(1995)]{Johnston1995} Johnston, K. V.,
  Spergel, D. N., \& Hernquist, L. 1995, \apj, 451, 598

\bibitem[Kinman et al.(2007)]{Kinman2007} Kinman, T.~D., Cacciari, C.,
  Bragaglia, A., Buzzoni, A., \& Spagna, A.\ 2007, \mnras, 375, 1381
 
\bibitem[Kotoneva et al.(2005)]{Kotoneva2005} Kotoneva, E., Innanen,
  K., Dawson, P.~C., Wood, P.~R., \& De Robertis, M.~M.\ 2005, \aap,
  438, 957
 
\bibitem[Law et al.(2003)]{Law2003} Law, D.~R., Majewski, S.~R.,
  Skrutskie, M.~F., Carpenter, J.~M., \& Ayub, H.~F.\ 2003, \aj, 126,
  1871

\bibitem[Law et al.(2005)]{Law2005} Law, D.~R., Johnston, K.~V., \&
  Majewski, S.~R.\ 2005, \apj, 619, 807

\bibitem[Lee et al.(1999)]{Lee1999} Lee, Y.-W., Joo, J.-M., Sohn,
  Y.-J., Rey, S.-C., Lee, H.-C., \& Walker, A.~R.\ 1999, \nat, 402, 55

\bibitem[Leon et al.(2000)]{Leon2000} Leon, S., Meylan, G., \& Combes,
  F.\ 2000, \aap, 359, 907
   
\bibitem[Majewski(1992)]{Majewski1992} Majewski, S.~R.\ 1992, \apjs,
  78, 87
 
\bibitem[Majewski(1993)]{Majewski1993} Majewski, S.~R.\ 1993, \araa,
  31, 575

\bibitem[Majewski(2004)]{Majewski2004} Majewski, S.~R.\ 2004, PASA,
  21, 197
 
\bibitem[Majewski et al.(2004)]{Majewskietal2004} Majewski, S.~R., et
  al.\ 2004, \aj, 128, 245
 
\bibitem[Majewski et al.(1994)]{Majewskietal1994} Majewski, S.~R.,
  Munn, J.~A., \& Hawley, S.~L.\ 1994, \apjl, 427, L37
 
\bibitem[Majewski et al.(1996)]{Majewskietal1996} Majewski, S.~R.,
  Munn, J.~A., \& Hawley, S.~L.\ 1996, \apjl, 459, L73
 
\bibitem[Majewski et al.(2000a)]{PaperI} Majewski, S.~R., Ostheimer,
  J.~C., Kunkel, W.~E., \& Patterson, R.~J.\ 2000a, \aj, 120, 2550

\bibitem[Majewski et al.(2000b)]{Majewski2000b} Majewski, S.~R.,
  Patterson, R.~J., Dinescu, D.~I., Johnson, W.~Y., Ostheimer, J.~C.,
  Kunkel, W.~E., \& Palma, C.\ 2000b, Liege International
  Astrophysical Colloquia, 35, 619

\bibitem[McCarthy et al.(1993)]{McCarthy1993} McCarthy, J.~K.,
  Sandiford, B.~A., Boyd, D., Booth, J. 1993, \pasp, 105, 881

\bibitem[Meza et al.(2005)]{Meza2005} Meza, A., Navarro, J.~F., Abadi,
  M.~G., \& Steinmetz, M.\ 2005, \mnras, 359, 93
 
\bibitem[Mizutani et al.(2003)]{Mizutani2003} Mizutani, A., Chiba, M.,
  \& Sakamoto, T.\ 2003, \apjl, 589, L89

\bibitem[Norris \& Da Costa(1995)]{Norris1995} Norris, J.~E. \& Da
  Costa, G.~S. 1995, \apj, 447, 680
 
\bibitem[Patterson et al.(2001)]{Patterson2001} Patterson, R.~J., et
  al.\ 2001, ASP Conf.~Ser.~246: IAU Colloq.~183: Small Telescope
  Astronomy on Global Scales, 246, 65
 
\bibitem[Proust \& Foy(1988)]{Proust1988} Proust, D. \& Foy, R.\ 1988,
  \apss, 145, 61

\bibitem[Reddy et al.(2003)]{Reddy2003} Reddy, B.~E., Tomkin, J.,
  Lambert, D.~L., Allende Prieto, C.  2003, \mnras, 340, 304

\bibitem[Reddy et al.(2006)]{Reddy2006} Reddy, B.~E., Lambert, D.~L.,
  Allende Prieto, C. 2006, \mnras, 367, 1329

\bibitem[Reid et al.(2009)]{Reid2009} Reid, M.~J., et al.\ 2009, \apj,
  700, 137
 
\bibitem[Searle \& Zinn(1978)]{Searle1978} Searle, L. \& Zinn,
  R.\ 1978, \apj, 225, 357

\bibitem[Smith \& Norris(1982)]{Smith1982} Smith, G. H. \& Norris,
  J. 1982, \apj, 254, 149
   
\bibitem[Smith et al.(2000)]{Smith2000} Smith, V.~V., Suntzeff, N.~B.,
  Cunha, K., Gallino, R., Busso, M., Lambert, D.~L., Straniero,
  O. 2000, \aj, 119, 1239

\bibitem[Stanford et al.(2006)]{Stanford2006} Stanford, L. M., Da
  Costa, G. S., Norris, J. E., \& Cannon, R. D. 2006, \apj, 647, 1075
 
\bibitem[Tsuchiya et al.(2003)]{Tsuchiya2003} Tsuchiya, T., Dinescu,
  D.~I., \& Korchagin, V.~I.\ 2003, \apjl, 589, L29
 
\bibitem[Tsuchiya et al.(2004)]{Tsuchiya2004} Tsuchiya, T., Korchagin,
  V.~I., \& Dinescu, D.~I.\ 2004, \mnras, 350, 1141

\bibitem[Wylie-de Boer et al.(2010)]{Wylie2010} Wylie-de Boer, E.,
  Freeman, K., \& Williams, M.\ 2010, \aj, 139, 636

\bibitem[Zacharias et al.(2010)]{Zacharias2010} Zacharias, N., et
  al.\ 2010, \aj, 139, 2184
 
\end{thebibliography}
\end{document}